# Enter Sandbox: Android Sandbox Comparison

Sebastian Neuner*, Victor van der Veen†, Martina Lindorfer‡, Markus Huber*,
Georg Merzdovnik*, Martin Mulazzani* and Edgar Weippl*

*SBA Research
Email: {sneuner,mhuber,gmerzdovnik,mmulazzani,eweippl}@sba-research.org
†The Network Institute, VU University Amsterdam; ITQ
Email: v.vander.veen@vu.nl
‡Secure Systems Lab, Vienna University of Technology
Email: mlindorfer@iseclab.org

*Abstract*—Expecting the shipment of 1 billion Android devices in 2017, cyber criminals have naturally extended their vicious activities towards Google's mobile operating system. With an estimated number of 700 new Android applications released every day, keeping control over malware is an increasingly challenging task. In recent years, a vast number of static and dynamic code analysis platforms for analyzing Android applications and making decision regarding their maliciousness have been introduced in academia and in the commercial world. These platforms differ heavily in terms of feature support and application properties being analyzed. In this paper, we give an overview of the state-of-the-art dynamic code analysis platforms for Android and evaluate their effectiveness with samples from known malware corpora as well as known Android bugs like Master Key. Our results indicate a low level of diversity in analysis platforms resulting from code reuse that leaves the evaluated systems vulnerable to evasion. Furthermore the Master Key bugs could be exploited by malware to hide malicious behavior from the sandboxes.

## I. INTRODUCTION

With an estimated market share of 70% to 80%, Android has become the most popular operating system for smartphones and tablets [1, 2]. Expecting a shipment of 1 billion Android devices in 2017 and with over 50 billion total app downloads since the first Android phone was released in 2008, cyber criminals naturally expanded their vicious activities towards Google's mobile platform. In the summer of 2012, the sophisticated *Eurograbber* attack showed that mobile malware may be a very lucrative business by stealing an estimated EUR 36,000,000 from bank customers in Italy, Germany, Spain and the Netherlands [3]. Also in 2012 the mobile malware *SMSZombie* [4] infected over 500,000 devices. In addition to obtaining administrator privileges, SMSZombie also gained control over the mobile SMS payment system of China Mobile, enabling the malware authors to secretly authorize payments.

An additional incentive for mobile malware authors to target Android instead of other mobile platforms is Android's open design that allows users to install applications from a variety of sources. However, the diversity of third-party app stores and volume of apps published poses a considerable challenge to security researchers and app store administrators when trying to identify malicious applications. With over 1 million apps available for download via the official Google Play Store [5] alone, and possibly another million spread among third-party app stores, it is possible to estimate that over 20,000 new applications are being released every month [6]. This requires scalable solutions for quickly analyzing new apps in order to isolate malicious and other possibly unwanted apps. Google reacted to the growing interest of miscreants in Android by introducing *Bouncer* in February 2012, a dynamic analysis sandbox that automatically checks apps submitted to the Google Play Store for malware [7]. However, research has shown that Bouncer's detection rate is still fairly low and that it can be easily bypassed [8, 9].

More than a year before Google's deployment of *Bouncer*, Bläsing et al. [10] were the first to present a dynamic analysis platform for Android applications called *AASandbox* (Android Application Sandbox). Since then, an ever increasing number of dynamic sandboxes for analyzing potentially malicious Android applications have been introduced in academia and in the commercial world. Similar to dynamic malware analysis platforms for Windows binaries such as CWSandbox [11] or Anubis [12], dynamic sandboxes for Android run targeted applications in a controlled environment to obtain a behavioral footprint. Results are normally presented to the user in the form of a report, possibly including a classification that indicates whether the app is benign or malicious. Most of these systems use some form of hybrid analysis, i.e., leveraging additional static analysis during a preprocessing phase to enhance the dynamic analysis results.

Egele et al. [13] systemized the existing knowledge on dynamic analysis of traditional malware. However, no such survey on dynamic malware analysis in a mobile context exists. In this paper, we systematically analyze the state-of-the-art dynamic analysis approaches for the Android platform and compare existing approaches in terms of provided features and analysis effectiveness.

In particular, the contributions of this paper are as follows:

- We survey current state-of-the-art Android malware detection techniques.

- We discuss methods to detect and fingerprint dynamic analysis sandboxes.

- We compare 16 dynamic analysis platforms for Android regarding their features, level of introspection, functionality and interdependencies.

- We evaluate the effectiveness of ten of these dynamic sandboxes using a selected set of malware samples from publicly available malware corpora. Furthermore, we analyze the susceptibility of these dynamic sandboxes to the so-called Master Key vulnerabilities.

The remainder of this paper is organized as follows: Section II gives an overview of current Android malware behavior, as well as distribution techniques. Section III discusses Android analysis techniques and numerous existing malware analysis frameworks. We then discuss our evaluation criteria and sandbox interdependencies in Section IV. Finally, we compare the sandboxes and evaluate their effectiveness and limitations in Section V, before we conclude in Section VI.

## II. ANDROID MALWARE

In order to gain a better understanding of the requirements for Android malware detection techniques to successfully analyze mobile malware samples, we give a brief overview on the current mobile malware threat landscape. In the following we discuss the motivation of mobile malware authors, different distribution methods as well as available malware data sets.

### A. Motivation

In August 2010 the first malicious Android application named *AndroidOS.FakePlayer* was discovered in the wild [14]. FakePlayer monetized infected devices by secretly sending text messages to premium numbers. Since then, both the sophistication as well as the amount of observed malware samples increased steadily. Recent reports focusing on mobile malware trends estimate that the number of malicious Android apps now ranges from 120,000 to 718,000 [15, 16].

The main motivation for mobile malware authors is *financial gain*. One way to monetize infected devices, leveraged by FakePlayer and countless other malware families since then, is by sending text messages to premium numbers, registered to malware authors. In addition to these so called *toll fraud* schemes, malware authors leverage their apps to spy on users and collect personal information. Mobile spyware has capabilities to forward private data to a remote server under the control of malware authors. In a more complex form, the malware could also receive commands from the server to start specific activities and become part of a *botnet*. Furthermore, mobile versions of the banking Trojan ZeuS (ZeuS-in-the-Mobile, or ZitMo) are a way to circumvent the two-factor authentication of online banking systems by stealing mobile TAN codes [17]. Broadcast receivers are an Android-specific feature of particular interest to malware authors as they can be used to launch a background service as soon as the device is started and secretly intercept and forward incoming text messages to a remote server. This capability was used in the aforementioned Eurograbber attack [3] in order to authorize financial transactions.

### B. Distribution

To lure victims into installing malicious apps, a common strategy employed by malware authors is to *repackage* popular applications with malicious payloads. The target applications often include paid applications, which are then offered for "free". Attackers commonly use third-party marketplaces to distribute their repackaged applications, as these marketplaces fail to verify submitted apps. Juniper Networks, for instance, found that malicious applications often originate from third-party marketplaces, with China and Russia being the world's leading suppliers [18]. Zhou et al. [19] analyzed repackaged apps in six alternative Android marketplaces and found that in addition to repackaging apps with malicious payloads, repackaged apps furthermore modify embedded advertising libraries to steal the ad-revenue of application developers.

Attackers can also leverage *drive-by downloads* adapted to the mobile context. Methods to trick users into installing malicious apps include in-app advertising [20], specially crafted websites or QR codes [21]. Drive-by downloads might however also exploit *platform-level vulnerabilities* to install malware in a stealth fashion. Due to Android's loose management to device software, Android versions have become fragmented, with only 1.8% of all devices running the latest Android version 4.4 (codename KitKat) [22], as of February 2014. This fragmentation makes new security features, as well as bugfixes for core components preventing against arbitrary code execution exploits, only available to a small group of users. Android versions prior to 2.3.7 are especially vulnerable to root exploits (examples include *RageAgainstTheCage* [23], *Exploid* [24] and *zergRush* [25]). While these exploits were originally developed to overcome limitations that carriers and hardware manufactures put on some devices, they have also been used by malware authors to obtain a higher privilege level without a user's consent. This approach allows malware to request only a few permissions during app installation, but still gaining root access to the entire system once the app is executed.

### C. Malware Data Sets

Access to known Android malware samples is mainly provided via the Android Malware Genome Project [26], Contagio Mobile [27] and VirusShare [28]. The Android Malware Genome Project contains over 1,200 Android malware samples from 49 families, collected from August 2010 to October 2011. Contagio Mobile offers an upload dropbox to share mobile malware samples among security researchers and currently hosts 164 archives, where some archives contain more than 16,000 samples. VirusShare also hosts a repository of malware samples with over 11,000 Android samples available to researchers. Furthermore, the multi-engine antivirus scanning services VirusTotal [29] and AndroTotal [30] provide researchers with access to submitted samples.

## III. ANDROID MALWARE ANALYSIS FRAMEWORKS

Ever since the first Android phones were released in 2008, researchers have proposed dozens of frameworks for a variety of purposes. In this section, we outline our efforts in systematizing these proposals by enumerating and analyzing this existing knowledge.

Analysis frameworks may use a number of techniques to produce a report about an app's functionality or perform a classification whether an app is benign or malicious. *Static analysis* techniques extract features from the Android application package (APK) and the Dalvik bytecode. *Dynamic*

*analysis* techniques monitor an app's behavior during runtime in a controlled environment. Results from static analysis can also enhance dynamic analysis, e.g. to efficiently stimulate a targeted application and trigger additional behavior, resulting in *hybrid analysis* approaches.

In the following paragraphs we briefly explain different analysis methods for both static and dynamic analysis and present available tools and frameworks for both approaches. Our evaluation mainly focuses on dynamic analysis frameworks, however, we also present various static analysis tools that might assist dynamic analysis and therefore be integrated in the presented dynamic analysis frameworks. For each dynamic analysis framework, we distinguish a number of characteristics and provide a brief summary. Based on their main purpose and approach, we classify the different research efforts into distinct categories.

### A. Static Analysis Tools

Static analysis tools may fall in one of the following categories:

- **Extraction of meta information:** Tools that extract information from an application's manifest and provide information about requested permissions, activities, services and registered broadcast receivers. Meta information is often used during later dynamic analysis in order to trigger an application's functionality.

- **Weaving:** Tools that rewrite bytecode of existing applications using a bytecode weaving technique. Using this technique, analysis frameworks can, for instance, insert tracing functionality into an existing application.

- **Decompiler:** Tools that implement a Dalvik bytecode decompiler or disassembler.

One of the most popular comprehensive static analysis tool for Android applications is *Androguard* [31]. It can disassemble and decompile Dalvik bytecode back to Java source code. Given two APK files, it can also compute a similarity value to detect repackaged apps or known malware. It also has modules that can parse and retrieve information from the app's manifest. Due to its flexibility, it is used by some other (dynamic) analysis frameworks that need to perform some form of static analysis. *APKinspector*[1] is a static analysis platform for Android application analysts and reverse engineers to visualize compiled Android packages and their corresponding DEX code. *Dexter*[2] is a web application designed for static analysis of Android applications. Its features are comparable with those of Androguard and APKinspector, however, it has some additional collaboration functionality to ease knowledge sharing among multiple researchers.

*APKtool*[3] is a tool for reverse engineering Android applications. It can decode resources to nearly original form and rebuild them into a new Android package after they have been modified. APKtool can be used to add additional features or extra support to existing applications without contacting the original author and thus assist bytecode weaving approaches. *Joe Sandbox Mobile* uses static instrumentation, which is equivalent to bytecode weaving in our classification. *Radare2*[4] is an open source reverse engineering framework which provides a set of tools to disassemble, debug, analyze, and manipulate binary Android files. Other disassembler tools for DEX files include *Dedexer*[5] and *smali/baksmali*[6]. Both read DEX files and convert them into an 'assembly-like' format which is largely influenced by the *Jasmin* syntax[7].

Other tools aim at enabling static analysis on Android applications by retargeting them to traditional `.class` files, which then can be processed by existing Java tools. One example is *ded* [32, 33], which later evolved into *Dare*[8]. Similarly, *dex2jar*[9] can convert an APK file directly to a `.jar` file and vice versa. *JEB*[10] is a commercial flexible interactive Android decompiler. It claims to be able to directly decompile Dalvik bytecode to Java source code, as well as disassemble an APK's contents so that users can view the decompressed manifest, resources, certificates, etc.

### B. Dynamic/Hybrid Analysis Frameworks

Dynamic analysis frameworks monitor the behavior of unknown applications at runtime by executing the targeted application in a controlled environment to generate a behavioral footprint. Dynamic analysis can monitor an app's behavior utilizing one (or more) of the following techniques:

- **Taint tracking:** Taint tracking tools are often used in dynamic analysis frameworks to implement system-wide dynamic taint propagation in order to detect potential misuse of users' private information.

- **Virtual machine introspection (VMI):** VMI-based frameworks [34] intercept events that occur within the emulated environment. *Dalvik VMI* based systems monitor the execution of Android APIs through modifications in the Dalvik VM. *Qemu VMI* based systems are implemented on the emulator level to enable the analysis of native code. However, emulators are prone to emulator evasion [35].

- **System call monitoring:** Frameworks can collect an overview of executed system calls, by using, for instance, VMI, `strace` or a kernel module. This enables (partial) tracing of native code.

- **Method tracing:** Frameworks can trace Java method invocations of an app in the Dalvik VM.

The first dynamic Android analysis framework developed was AASandbox [10], presented in October 2010. It implements a system call monitoring approach using a loadable kernel module. Furthermore, it uses the resulting system call footprint to discover possibly malicious applications.

A popular taint tracking framework is TaintDroid [36]. TaintDroid is implemented upon the Dalvik VM and monitors applications for the leakage of sensitive information. However, ScrubDroid [37] presented a number of attacks to circumvent dynamic taint analysis. VetDroid [38] is a dynamic framework that measures actual permission use behavior by dynamically building a permission use graph. Their data tainting method is built upon TaintDroid, but improves it by identifying implicit and explicit permission use points.

---

[1] https://github.com/honeynet/apkinspector/
[2] http://dexter.dexlabs.org/
[3] https://code.google.com/p/android-apktool/
[4] http://radare.org/y/
[5] http://dedexer.sourceforge.net
[6] https://code.google.com/p/smali/
[7] Jasmin is an assembler for the Java VM: http://jasmin.sourceforge.net
[8] http://siis.cse.psu.edu/dare/
[9] http://code.google.com/p/dex2jar/
[10] http://www.android-decompiler.com/

DroidBox [39] uses TaintDroid to detect privacy leaks, but also comes with a Dalvik VM patch to monitor the Android API and report file system and network activity, the use of cryptographic operations and cell phone usage such as sending SMS and making phone calls. However, it can be easily bypassed if applications include their own libraries, in particular by using it for cryptographic operations. In the latest version, DroidBox utilizes bytecode weaving to insert monitoring code into the app under analysis[11]. ANANAS [40] is a recently published framework focusing on a modularized architecture with analysis modules for logging e.g. file system activity, network activity or system calls. It also employs API monitoring similar to the most recent version of DroidBox and uses a loadable kernel module for system call monitoring.

DroidScope [41] uses VMI to reconstruct Dalvik and native code instruction traces. The authors also implemented their own data tainting method, named TaintTracker. One of the huge benefits is that it does not require any changes to the Android sources, however JIT has to be selectively disabled as it blurs the Dalvik instruction boundaries. Andrubis [42] leverages various techniques, is VMI-based and monitors events in the Dalvik VM as well as native code for system-calls through QEMU VMI. It further employs TaintDroid for data tainting. TraceDroid [43] has the benefit of generating a complete method trace output and was also integrated into Andrubis. It further captures network traffic and refrains from analyzing native code, as `ltrace` and `strace` are deemed sufficient for this purpose. This is also the approach used by Mobile Sandbox [44], which runs the app in the emulator and uses `ltrace` to track native code. Finally, CopperDroid [45] uses a VMI-based approach as well, but compared to the other frameworks it is also capable of analyzing IPC and RPC-based communication between applications.

Andrubis, TraceDroid, Mobile Sandbox and CopperDroid all perform input stimulation, most notably the monkey exerciser and the emulation of common events like GPS lock, SMS received or boot-completed. To further enhance code coverage, several frameworks for automatically extracting and exercising UI based triggers have been proposed. The SmartDroid framework [46] statically extracts the function call graph and activity call graph, and then dynamically traverses these graphs to find elements that trigger sensitive behavior. In contrast to other frameworks, however, it does not focus on possibly malicious activities like sending SMS or accessing files. AppsPlayground [47] is a framework that automates analysis of Android applications and monitors taint propagation (using TaintDroid), specific API calls and system calls. Its main contribution is a heuristic-based intelligent black-box execution approach to explore the app's GUI. The goal of AppIntent [48] is to distinguish user-intended data transmission from unintended ones, and is as such related to SmartDroid. It uses an event-space constraint guided symbolic execution technique to construct event sequences, which effectively reduces the event search space in symbolic execution for Android apps.

A number of additional dynamic analysis platforms have been implemented and made available to the public via web applications: SandDroid[12], VisualThreat[13], ForeSafe[14] and the Joe Sandbox Mobile APK Analyzer[15]. These frameworks, however, come often with very little public documentation on how they operate, which makes it hard to make statements on any new approaches used by these implementations. It is likely that most of these platforms use (modified versions of) existing tools like DroidBox, TaintDroid and Androguard to complement their dynamic analysis engine. This is confirmed for example on SandDroid's webpage, which states that it is powered by both DroidBox and Androguard. While we could not find a public reference which Android version is used by Joe Sandbox Mobile, it was possible to create APK files with varying API level requirements and fingerprint the Android version to API level 15. This either is Android version 4.0.3 or 4.0.4, which is an upper bound of supported Android versions. Unfortunately, this fingerprinting technique was unsuccessful with ForeSafe and VisualThreat.

### C. Miscellaneous Frameworks

For the sake of completeness it is also worth mentioning various other frameworks that offer online services for the analysis of Android apps based on their static features or meta information: Badger[16] provides a web service for extracting required permissions as well as included advertising applications from uploaded APK files. Undroid[17] performs slightly more comprehensive static analysis and extracts requested permissions as well as API calls that use these permissions. The Android Observatory [49] also extracts required permissions from the manifest and further tries to match the app's certificate against those of known malware.

Finally, VirusTotal [29] and AndroTotal [30] are two multi-engine anti-virus scanning services. VirusTotal uses static AV-based scanner engines to assess the maliciousness of a submitted APK file. AndroTotal uses dynamic sandboxing to test samples against mobile malware detectors, similar to VirusTotal. However, it uses the apps themselves on physical devices as well as in the emulator instead of command line versions with the same underlying signature database. It also provides links to analysis results from other frameworks: VirusTotal, CopperDroid, ForSafe, SandDroid and Andrubis.

### IV. EVALUATION CRITERIA

We present an overview of all dynamic analysis frameworks discussed so far in Table I and list, whether the source code is publicly available (src) or the service is available through a web interface (www).

---

[11] http://code.google.com/p/droidbox/wiki/APIMonitor
[12] http://sanddroid.xjtu.edu.cn
[13] http://www.visualthreat.com
[14] http://www.foresafe.com
[15] http://www.apk-analyzer.net
[16] http://davidson-www.cs.wisc.edu/baa
[17] http://www.av-comparatives.org/avc-analyzer/

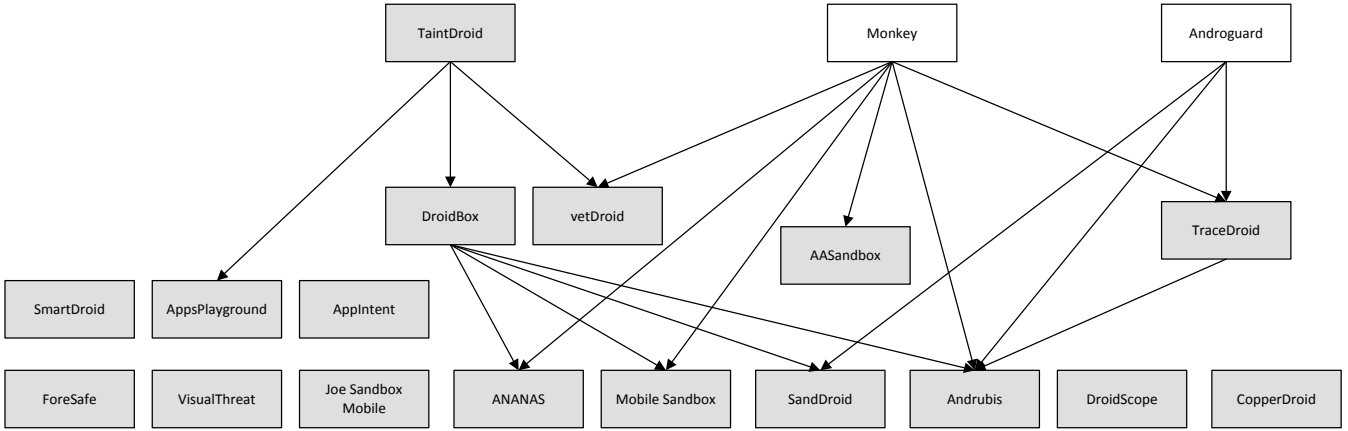

Fig. 1. Interdependency graph illustrating sandboxes relying on other tools or services.

TABLE I. OVERVIEW OF ANDROID ANALYSIS FRAMEWORKS AND THEIR AVAILABILITY - EITHER AS SOURCE CODE (SRC) OR THROUGH A PUBLIC WEB INTERFACE (WWW).

| Framework | src | www | Framework | src | www |
|---|---|---|---|---|---|
| AASandbox [10] | | | ForeSafe | | • |
| AppIntent [48] | | | Joe Sandbox Mobile | | • |
| ANANAS [40] | | | Mobile Sandbox [44] | | • |
| AndroTotal [30] | | • | SandDroid | | • |
| Andrubis [42] | | • | SmartDroid [46] | | |
| AppsPlayground [47] | • | | TaintDroid [36] | • | |
| CopperDroid [45] | | • | TraceDroid [43] | | • |
| DroidBox [39] | • | | vetDroid [38] | | |
| DroidScope [41] | • | | VisualThreat | | • |

### A. Features of Interest

In order to compare the discussed dynamic analysis frameworks we manually examined them and compiled a feature list for further evaluation. First of all, we are interested in implementation details such as the *Android version* that is supported by the sandbox. Android is under active development, and at the time of writing Android 4.4 (codename KitKat) is the most recent version. We are also interested in the *inspection level* and which methods the sandboxes use to capture the dynamic behavior of an app. As discussed in Section III-B, multiple methods are suitable: Modifying stock Android for additional logging capabilities is one convenient way to capture app behavior, in particular the kernel or the APIs. ANANAS for example uses a loadable kernel module for its analysis. The downside of this approach is that the kernel patches cannot be easily ported to other versions of Android, as the internals can change between updates. Secondly, as Android programs are executed within the Dalvik VM, some approaches modify the Dalvik VM to capture the dynamic behavior of an app. TaintDroid and DroidBox use this method for example. Downside of this approach is that only Java calls are traceable and native code is not captured. Lastly, it is possible to modify QEMU for enhanced logging mechanisms and to use true virtual machine introspection. This is a holistic method, as it works on a lower level compared to the other methods. However, arm-to-x86 emulation is known to be slow, and as such it has a high overhead. Some sandboxes even use combinations of various methods e.g., Andrubis uses virtual machine introspection through QEMU in combination with a modified Dalvik VM.

Additionally, some sandboxes use *taint tracking* to detect if the users private information is leaked. Furthermore, sandboxes usually perform some form of *static analysis* such as parsing the requested permissions from the manifest. *GUI interactions* are usually employed to simulate user activity and enhance the coverage of dynamic analysis. We can further distinguish the sandboxes by the features they analyze. *Network activity* as well as *file activity* is of particular interest in malware analysis, as malware usually reads and writes local files and relies on obtaining commands from a command and control server in case of a botnet [50]. As discussed in Section II-A, malware is often used to generate revenue for the malware creator by sending premium SMS or calling premium phone numbers causing financial damage to the victim. As such, we are interested if the sandbox is able to capture *phone activity*. Furthermore, we are interested if the analysis frameworks are able to analyze *native code*.

### B. Framework Interdependency

Many sandboxes build on previous work and incorporate existing tools in their service. In particular a combination of the open source solutions TaintDroid and DroidBox for dynamic analysis and Androguard for static analysis are heavily used. As we will show in the evaluation in Section V, this can be a problem if the malware is capable of evading these underlying analysis tools. Furthermore, many sandboxes rely on the Application Exerciser Monkey[18] (or Monkey) to simulate user input, even though the intended use case of Monkey was to stress test applications with pseudorandom user events. Figure 1 shows the full interdependency graph depicting which sandboxes rely on which other tools or services.

### C. Sandbox Fingerprinting

Dynamic sandboxes are often rather easy to fingerprint, which is a problem that has been extensively studied for PC malware [51, 52, 53]. First of all, system emulators like

---

[18]http://developer.android.com/tools/help/monkey.html

QEMU [35, 54] and virtual machines [55] are detectable. With the numerous sensors and interfaces of a smartphone, a real device is even more challenging to simulate than PC hardware. GPS, screen size, and the motion sensor are just a few examples of what has to be considered when building sandboxes for mobile devices. Malware can also simply wait for a certain amount of time, as dynamic sandboxes cannot execute an app indefinitely [56] due to limited resources. Other methods to detect if the app is running within a sandbox include checking the application's signature. This makes it possible to detect platforms like the recent version of DroidBox that modify the app's bytecode.

Fingerprinting methods can also rely on implementation details of specific sandboxes. For example, AASandbox uses 500 monkey events to simulate user input, with 1 second between events [10]. While this is of course just a configuration parameter, malware could easily detect AASandbox if these parameters are not modified. TraceDroid on the other hand initializes the Monkey with the same seed value for the pseudo-random number generator (currently set to 1337) [43]. This is intended to ensure that re-runs of the exerciser generate the same sequence of events, but this is however also detectable by malware. SmartDroid [46] waits for 10 seconds before starting UI interactions during dynamic analysis. There is yet no ground truth on the distribution and statistical likelihood of user events.

## V. Evaluation Results

We evaluated a total of 10 dynamic sandboxes based on the feature set outlined above, skipping sandboxes that are not available to us in any form. Our analysis is based on the academic publications as well as available documentation. Table II lists the extracted features for all sandboxes. We furthermore tested the sandboxes with selected malware samples, which we discuss in the following paragraphs.

### A. Malware Samples

For our evaluation we used real-world malware samples from four different families or categories. For each of those families we used two different samples. All of the samples have been publicly analyzed and described, either as part of the Android Malware Genome Project [26] or in blogposts or technical reports from antivirus vendors. All four families were chosen due to either feature coverage (regarding [26] coverage in terms of financial charges and personal information stealing) or due to interesting behaviour like sophisticated privilege escalation techniques or evasion of specific sandboxes.

*a) Obad:* In 2013, Kaspersky Lab [57] described Obad as one of the most sophisticated mobile malware to date. The user has to install the malware manually with 24 different permissions, which grant nearly total access to the device. Obad is able to send SMS, send data to an URL, download and install files, and transfer files via bluetooth. Additionally, the application tries to connect to a command and control server and is able to join a botnet. Malware of the Obad family tries to evade detection from several sandboxes with anti-emulation and anti-decompilation techniques, like checking for the *Android.os.build.MODEL* value to quit the applications execution if the default value of the emulator is present [58, 59].

*b) Geinimi:* The second Android malware family we selected for our analysis is Geinimi, which is part of the Android Malware Genome Project. We selected Geinimi due to its various malicious payloads: remote control over the Internet, starting phone calls, sending SMS as well as leaking sensitive data stored on the phone.

*c) DroidKungFu:* DroidKungFu [60] is part of the Android Malware Genome Project as well, and is a malware family that can perform various forms of privilege escalation. It collects various phone related data such as the IMEI number, the phone model etc. and sends it to a server. This malware family is of particular interest since it is able to execute various exploits, for example privilege escalation using the RageAgainstTheCage exploit [23]. This exploit decrypts itself during runtime to exploit the `adb` resource exhaustion bug for root access. After this procedure the application is able gain root privileges and hide itself, drop other malware, etc.

*d) Basebridge/Nyleaker:* The fourth malware category we used for our evaluation are the malware samples Basebridge and Nyleaker. Both samples are able to successfully evade analysis by abusing the Androguard tool (which is used by many sandboxes) by presenting itself as an invalid application by either using a corrupt APK or an invalid manifest file within the APK. Their malicious behavior includes sending SMS to premium services, executing local privilege escalation exploits and leaking personal information.

*e) Master Key:* The last evaluated category is the Master Key set that exploits several Android vulnerabilities related to the handling of ZIP files[19].

- *Bug 8219321*[20]: The ZIP format allows multiple occurrences of the same filename within one archive. This can lead to a serious vulnerability, if the implementation of the unpacking of the archive differs in parts of the systems: The Android signature verifier checks the first file for integrity and compares it with the META-INF directory. The Android installer however uses the second file for installation. Symantec reported that this bug was already exploited by malware in the wild [61]. Specifically malware samples added an additional malicious classes.dex as well as a second manifest file to a benign app without breaking its signature.

- *Bug 9695860*[21]: This bug is a simple signed-unsigned integer mismatch between different parts of the Android code. The ZIP file header contains the fields for filename length and extra field length. The signature verification code treats the contents of these fields as signed 16-bit integers, which converts the input of "0xFFFD" to "-3". Thus, the Android verifier jumps three bytes backwards, instead of forward and therefore skips the extra field by reading the part "dex" of "classes.dex". The application loading code however treats the fields as unsigned 16-bit integers,

---
[19]APK files are ZIP compressed files, based on the JAR format.
[20]http://nakedsecurity.sophos.com/2013/07/10/anatomy-of-a-security-hole-googles-android-master-key-debacle-explained
[21]http://nakedsecurity.sophos.com/2013/07/17/anatomy-of-another-android-hole-chinese-researchers-claim-new-code-verification-bypass/

TABLE II. COMPARISON OF ANDROID MALWARE ANALYSIS SANDBOXES.

| Framework | Implementation Details | | Analysis Type | | | Analyzed Features | | | |
|---|---|---|---|---|---|---|---|---|---|
| | Android Version | Inspection Level | Static | Tainting | GUI Interactions | File | Network | Phone | Native Code |
| *AASandbox* | — | Kernel | ● | | ● | ● | ● | ● | |
| *AppIntent* | 2.3 | Kernel | | ● | ● | ● | ● | ● | |
| *ANANAS* | 2.3-4.2 | Kernel | ● | | ● | ● | ● | ● | ● |
| *Andrubis* | 2.3.4 | QEMU & Dalvik | ● | ● | ● | ● | ● | ● | ● |
| *AppsPlayground* | — | Kernel | ● | ● | ● | ● | ● | ● | |
| *CopperDroid* | 2.2.3 | QEMU | ● | | ● | ● | ● | ● | ● |
| *DroidBox* | 2.3-4.1 | Kernel | | ● | | ● | ● | ● | |
| *DroidScope* | 2.3 | Kernel & Dalvik | | ● | | ● | ● | ● | ● |
| *ForeSafe* | ? | ? | ● | | ● | ● | ● | ● | |
| *Joe Sandbox Mobile* | 4.0.3 / 4.0.4 | Static Instrumentation | ● | | ● | ● | ● | ● | ● |
| *Mobile Sandbox* | 2.3.4 | Dalvik | ● | ● | ● | ● | ● | ● | ● |
| *SandDroid* | ? | ? | ● | ● | ? | ● | ● | ? | ? |
| *SmartDroid* | 2.3.3 | Kernel | ● | ● | ● | ● | ● | ● | |
| *TraceDroid* | 2.3.4 | Dalvik | ● | | ● | ● | ● | ● | |
| *vetDroid* | 2.3 | Kernel & Dalvik | ● | ● | ● | ● | ● | ● | |
| *VisualThreat* | ? | ? | ● | | | | | | ● |

which could match the start of a second uncompressed dex file supplied in the file data section.

- *Bug 9950697*[22]: This last version of the Master Key vulnerability is caused by the redundant storage of the "filename length" field in the ZIP header. This field indicates how many bytes between the filename and the actual file data exist and also how many bytes in the central directory exist, to reach the next directory entry in the archive. It is possible to provide a real filename length for the verifier, that verifies the trustworthy original file data and fake a filename length in the local header for the loader, that later executes the malware code.

Additionally, a certain bug in Python[23] can be used to create APK files which are not processed correctly by certain sandboxes. This Python bug triggers an exception when the length field in the ZIP header is zero. While this is a Python issue and not related directly to Android, many tools and sandboxes which are based on Python are affected and malware authors could craft APK files that trigger this bug to evade analysis.

## B. Analysis Results

We submitted the 12 malware samples (8 from known corpora, and 4 exploiting one Master Key vulnerability each) to all sandboxes which were available at the time of writing and which use some form of dynamic analysis within a sandbox. Sandboxes which were not available during our evaluation were excluded, leaving ten sandboxes. Our evaluation is based on whether the sandboxes detect the malicious behavior. The results of our evaluation are outlined in Table III.

For every family of malware, we submitted two different samples (separated by a "/" in Table III). Andrubis was able to analyze every submitted sample, while the Nyleaker malware was classified as benign. ForeSafe was able to successfully flag every submitted sample as high risk malware. SandDroid was unable to analyze any submitted sample. As for CopperDroid, just one version of Geinimi and DroidKungFu were detected.

---

[22]http://nakedsecurity.sophos.com/2013/11/06/anatomy-of-a-file-format-problem-yet-another-code-verification-bypass-in-android/
[23]http://bugs.python.org/issue14315

TABLE III. ANALYSIS RESULTS OF ONLINE SANDBOXES FOR TWO SAMPLES PER MALWARE FAMILY ("●"=DETECTED, "○"=NOT DETECTED, "-"=ANALYSIS ERROR).

| Framework | Obad | Geinimi | DroidKungFu | Basebridge/ Nyleaker |
|---|---|---|---|---|
| *Andrubis* | ● / ● | ● / ● | ● / ● | ● / ○ |
| *CopperDroid* | - / - | ● / - | - / ● | - / - |
| *ForeSafe* | ● / ● | ● / ● | ● / ● | ● / ● |
| *Joe Sandbox Mobile* | ● / ● | ● / ● | ● / ● | ● / ● |
| *Mobile Sandbox* | - / - | - / - | - / - | - / - |
| *SandDroid* | - / - | - / - | - / - | - / - |
| *TraceDroid* | ● / ● | ● / ● | ● / ● | ● / ● |
| *VisualThreat* | ● / - | ● / ● | ● / ● | ● / ● |

TABLE IV. EVALUATION OF SANDBOXES WITH MASTER KEY SAMPLES: "●" INDICATES THAT THE SAMPLE WAS SUCCESSFULLY EXECUTED, "-" INDICATES, THAT THE SANDBOX WAS NOT ABLE TO EXECUTE THE SAMPLE.

| Framework | Bug 8219321 | Bug 9695860 | Bug 9950697 | Python ZIP Bug |
|---|---|---|---|---|
| *Andrubis* | ● | - | - | ● |
| *CopperDroid* | - | - | - | - |
| *ForeSafe* | ● | ● | ● | ● |
| *TraceDroid* | ● | - | - | ● |
| *VisualThreat* | ● | ● | - | ● |

For all other samples the analysis was aborted with an "Installation Error". Joe Sandbox Mobile as well as TraceDroid on the contrary were able to analyze every version of every sample we submitted. Joe Sandbox Mobile furthermore correctly flagged every sample as malicious in the generated reports.

We also compiled an APK for each of the described Master Key bugs and submitted them to our selection of sandboxes to see if they could analyze malicious APKs exploiting those vulnerabilities. We present the results in Table IV. As we will discuss in Section V-C we were not able to get results from all sandboxes for our evaluation in time, as several sandboxes were out of service or did not produce any reports in a reasonable amount of time.

## C. Limitations of Existing Sandboxes

We observed that many sandboxes were not able to fully analyze our samples, were prone to bugs or evasion techniques, or were either no longer maintained or not publicly available anymore at the time of writing. As some versions of the Obad family evade DroidBox, Mobile Sandbox was not able to analyze these. We verified this in a personal conversation with the author of Mobile Sandbox. The other samples, namely Geinimi, DroidKungFu, Basebridge and Nyleaker, are also marked as errors in our table as there are currently over 300,000 samples in the queue to be analyzed by the dynamic sandbox. Any submitted sample would thus be analyzed after approximately 400 days, which is too long in our point of view. Unfortunately, this number does not seem to decrease. SandDroid accepts samples for analysis but is not able to perform any analysis on the samples. After submitting the samples and waiting for about two weeks, the samples are either not in the report database or marked as still being analyzed by the framework.

Some frameworks mentioned in Table II were excluded from our evaluation: AASandbox, AppIntent, ANANAS, SmartDroid as well as vetDroid were neither available as source code nor as an online submission. We were not able to run any samples for analysis within these frameworks. Because AppsPlayground has no online submission possibility[24] it was out of scope and therefore also excluded from our evaluation.

*D. Discussion*

Every sandbox carries out a multitude of analysis techniques, ranging from basic static to rigorous dynamic analysis techniques. Due to the large number of available analysis frameworks, on- as well as offline, it is impossible for a user to determine which framework offers the most and comprehensive set of features. This fact is substantiated by the ongoing improvements in the mobile sector, like Google's new Android Runtime (ART) technology[25], that will also complicate static and dynamic analysis in the future due to possibly pre-compiled applications.

Supporting these facts, there is no Swiss-army-knife-sandbox that on one hand offers every possible feature and on the other hand is readily available. Nevertheless, also on mobile devices the cat-and-mouse-game between malware authors and security researchers is continuously ongoing, as it has been on personal computers for many years. One possibility to deal with that problem could be one non-commercially driven analysis framework that implements all static and dynamic analysis techniques discussed in the literature so far, and is fully maintained as well as continuously extended by researching volunteers. Until this goal is reached, malware authors will be ahead of defending researchers and industrial professionals.

We tried to contact every sandbox creator and notified them of our findings.

## VI. CONCLUSION

In this paper we conducted a comprehensive study on detection methods of Android malware. First, we provided an overview on current Android malware distribution techniques and the motivation of malware authors. Secondly, we analyzed available dynamic analysis platforms for Android and examined their interdependencies. An evaluation of ten sandboxes which are available as online services, using samples from four real-life Android malware families, showed that detection rates vary. In addition we found that popular dynamic sandboxes are susceptible to well known vulnerabilities like the Master Key vulnerabilities, which could be potentially misused by Android malware to thwart analysis. In conclusion, our findings show that while current dynamic sandboxes are valuable for academic research to analyze Android malware, they can not be considered an effective defense against malware on their own. Furthermore, interdependencies between platforms caused by code reuse can lead to challenges in detecting malware that targets specific platform limitations, as all analysis platforms which share the same code base are affected. Thus, while a number of different analysis sandboxes exist, diversity among them is low.

ACKNOWLEDGEMENTS

The research was funded by COMET K1, FFG - Austrian Research Promotion Agency. The research leading to these results has also received funding from the European Union Seventh Framework Programme (FP7/2007-2013) under grant agreement no 257007. Moreover this work has been carried out within the scope of u'smile, the Josef Ressel Center for User-Friendly Secure Mobile Environments. We gratefully acknowledge funding and support by the Christian Doppler Gesellschaft, A1 Telekom Austria AG, Drei-Banken-EDV GmbH, LG Nexera Business Solutions AG, and NXP Semiconductors Austria GmbH.

## REFERENCES

[1] Canalys, "Over 1 billion Android-based smart phones to ship in 2017," http://www.canalys.com/newsroom/over-1-billion-android-based-smart-phones-ship-2017, June 2013.
[2] R. Llamas, R. Reith, and M. Shirer, "Apple Cedes Market Share in Smartphone Operating System Market as Android Surges and Windows Phone Gains, According to IDC," http://www.idc.com/getdoc.jsp?containerId=prUS24257413, August 2013.
[3] E. Kalige and D. Burkey, "A Case Study of Eurograbber: How 36 Million Euros was Stolen via Malware," https://www.checkpoint.com/products/downloads/whitepapers/Eurograbber_White_Paper.pdf, December 2012.
[4] M. Lennon, "Resilient 'SMSZombie' Infects 500,000 Android Users in China," http://www.securityweek.com/resilient-smszombie-infects-500000-android-users-china, August 2012.
[5] C. Warren, "Google Play Hits 1 Million Apps," http://mashable.com/2013/07/24/google-play-1-million, July 2013.
[6] AppBrain, "Number of available Android applications," http://www.appbrain.com/stats/number-of-android-apps.
[7] H. Lockheimer, "Android and Security," http://googlemobile.blogspot.com/2012/02/android-and-security.html, February 2012.
[8] X. Jiang, "An Evaluation of the Application ("App") Verification Service in Android 4.2," http://www.cs.ncsu.edu/faculty/jiang/appverify, Dec. 2012.
[9] J. Oberheide and C. Miller, "Dissecting the Android Bouncer," in *SummerCon*, 2012.

---

[24]http://list.cs.northwestern.edu/mobile/
[25]http://source.android.com/devices/tech/dalvik/art.html

[10] T. Bläsing, L. Batyuk, A.-D. Schmidt, S. A. Camtepe, and S. Albayrak, "An Android Application Sandbox System for Suspicious Software Detection," in *Proceedings of the 5th International Conference on Malicious and Unwanted Software (MALWARE)*, 2010.

[11] C. Willems, T. Holz, and F. Freiling, "Toward Automated Dynamic Malware Analysis Using CWSandbox," *IEEE Security and Privacy*, vol. 5, no. 2, 2007.

[12] U. Bayer, C. Kruegel, and E. Kirda, "TTAnalyze: A Tool for Analyzing Malware," in *Proceedings of the 15th European Institute for Computer Antivirus Research (EICAR) Annual Conference*, 2006.

[13] M. Egele, T. Scholte, E. Kirda, and C. Kruegel, "A Survey on Automated Dynamic Malware Analysis Techniques and Tools," *ACM Computing Surveys Journal*, vol. 44, no. 2, 2012.

[14] D. Maslennikov, "First SMS Trojan for Android," https://www.securelist.com/en/blog/2254/First_SMS_Trojan_for_Android, August 2010.

[15] Alcatel-Lucent, "Kindsight Security Labs Malware Report - Q2 2013," http://www.kindsight.net/sites/default/files/Kindsight-Q2-2013-Malware-Report.pdf, Jul. 2013.

[16] Trend Micro, "TrendLabs 2Q 2013 Security Roundup," http://www.trendmicro.com/cloud-content/us/pdfs/security-intelligence/reports/rpt-2q-2013-trendlabs-security-roundup.pdf, August 2013.

[17] D. Maslennikov, "Zeus-in-the-Mobile — Facts and Theories," http://www.securelist.com/en/analysis/204792194, October 2011.

[18] Juniper Networks, "Juniper Networks Third Annual Mobile Threats Report," http://www.juniper.net/us/en/local/pdf/additional-resources/jnpr-2012-mobile-threats-report.pdf, June 2013.

[19] W. Zhou, Y. Zhou, X. Jiang, and P. Ning, "Detecting Repackaged Smartphone Applications in Third-Party Android Marketplaces," in *Proceedings of the 2nd ACM Conference on Data and Application Security and Privacy (CODASPY)*, 2012.

[20] A. Bibat, "GGTracker Malware Hides As Android Market," http://www.androidauthority.com/ggtracker-malware-hides-as-android-market-17281/, June 2011.

[21] H. Yao and D. Shin, "Towards Preventing QR Code Based Attacks on Android Phone Using Security Warnings," in *Proceedings of the 8th ACM Symposium on Information, Computer and Communications Security (ASIACCS)*, 2013.

[22] Google, "Android Developers: Platform Versions," http://developer.android.com/about/dashboards/index.html.

[23] "RageAgainstTheCage," http://thesnkchrmr.wordpress.com/2011/03/24/rageagainstthecage/, 2011.

[24] "udev Exploit (exploid)," http://thesnkchrmr.wordpress.com/2011/03/27/udev-exploit-exploid, 2011.

[25] "zergRush," http://github.com/revolutionary/zergRush.

[26] Y. Zhou and X. Jiang, "Dissecting Android Malware: Characterization and Evolution," in *Proceedings of the 33rd Annual IEEE Symposium on Security and Privacy (S&P)*, 2012.

[27] "Contagio Mobile," http://contagiominidump.blogspot.com.

[28] "VirusShare," http://www.virusshare.com.

[29] "VirusTotal," http://www.virustotal.com.

[30] F. Maggi, A. Valdi, and S. Zanero, "AndroTotal: A Flexible, Scalable Toolbox and Service for Testing Mobile Malware Detectors," in *Proceedings of the 3rd ACM Workshop on Security and Privacy in Smartphones and Mobile Devices (SPSM)*, 2013.

[31] A. Desnos and G. Gueguen, "Android: From Reversing to Decompilation," in *Black Hat Abu Dhabi*, Dec. 2011.

[32] W. Enck, D. Octeau, P. McDaniel, and S. Chaudhuri, "A Study of Android Application Security," in *Proceedings of the 20th USENIX Security Symposium*, 2011.

[33] D. Octeau, W. Enck, and P. McDaniel, "The ded Decompiler," Network and Security Research Center, Department of Computer Science and Engineering, Pennsylvania State University, University Park, PA, USA, Tech. Rep. NAS-TR-0140-2010, Sep. 2010.

[34] T. Garfinkel and M. Rosenblum, "A Virtual Machine Introspection Based Architecture for Intrusion Detection," in *Proceedings of the 10th Annual Network & Distributed System Security Symposium (NDSS)*, 2003.

[35] T. Raffetseder, C. Kruegel, and E. Kirda, "Detecting System Emulators," in *Proceedings of the 10th Information Security Conference (ISC)*, 2007.

[36] W. Enck, P. Gilbert, B.-G. Chunn, L. P. Cox, J. Jung, P. McDaniel, and A. N. Sheth, "TaintDroid: An Information-Flow Tracking System for Realtime Privacy Monitoring on Smartphones," in *Proceedings of the 9th USENIX Symposium on Operating Systems Design and Implementation (OSDI)*, 2010.

[37] G. Sarwar, O. Mehani, R. Boreli, and M. A. Kaafar, "On the Effectiveness of Dynamic Taint Analysis for Protecting Against Private Information Leaks on Android-based Devices," in *Proceedings of the 10th International Conference on Security and Cryptography (SECRYPT)*, 2013.

[38] Y. Zhang, M. Yang, B. Xu, Z. Yang, G. Gu, P. Ning, X. Wang, and B. Zang, "Vetting Undesirable Behaviors in Android Apps with Permission Use Analysis," in *Proceedings of the 20th ACM Conference on Computer and Communications Security (CCS)*, 2013.

[39] "DroidBox: An Android Application Sandbox for Dynamic Analysis," https://code.google.com/p/droidbox/.

[40] T. Eder, M. Rodler, D. Vymazal, and M. Zeilinger, "ANANAS - A Framework For Analyzing Android Applications," in *Proceedings on the 1st International Workshop on Emerging Cyberthreats and Countermeasures (ECTCM)*, 2013.

[41] L. K. Yan and H. Yin, "DroidScope: Seamlessly Reconstructing the OS and Dalvik Semantic Views for Dynamic Android Malware Analysis," in *Proceedings of the 21st USENIX Security Symposium*, 2012.

[42] L. Weichselbaum, M. Neugschwandtner, M. Lindorfer, Y. Fratantonio, V. van der Veen, and C. Platzer, "Andrubis: Android Malware Under The Magnifying Glass," Vienna University of Technology, Tech. Rep. TR-ISECLAB-0414-001, 2014.

[43] V. van der Veen, "Dynamic Analysis of Android Malware," *Internet & Web Technology Master thesis, VU University Amsterdam*, 2013.

[44] M. Spreitzenbarth, F. Freiling, F. Echtler, T. Schreck, and J. Hoffmann, "Mobile-Sandbox: Having a Deeper Look into Android Applications," in *Proceedings of the 28th Annual ACM Symposium on Applied Computing (SAC)*, 2013.

[45] A. Reina, A. Fattori, and L. Cavallaro, "A System Call-Centric Analysis and Stimulation Technique to Automatically Reconstruct Android Malware Behaviors," in *Proceedings of the 6th European Workshop on System Security (EUROSEC)*, 2013.

[46] C. Zheng, S. Zhu, S. Dai, G. Gu, X. Gong, X. Han, and W. Zou, "SmartDroid: An Automatic System for Revealing UI-based Trigger Conditions in Android Applications," in *Proceedings of the 2nd ACM Workshop on Security and Privacy in Smartphones and Mobile Devices (SPSM)*, 2012.

[47] V. Rastogi, Y. Chen, and W. Enck, "AppsPlayground: Automatic Security Analysis of Smartphone Applications," in *Proceedings of the 3rd ACM Conference on Data and Application Security and Privacy (CODASPY)*, 2013.

[48] Z. Yang, M. Yang, Y. Zhang, G. Gu, P. Ning, and X. S. Wang, "AppIntent: Analyzing Sensitive Data Transmission in Android for Privacy Leakage Detection," in *Proceedings of the 20th ACM Conference on Computer and Communications Security (CCS)*, 2013.

[49] D. Barrera, J. Clark, D. McCarney, and P. C. van Oorschot, "Understanding and Improving App Installation Security Mechanisms Through Empirical Analysis of Android," in *Proceedings of the 2nd ACM Workshop on Security and Privacy in Smartphones and Mobile Devices (SPSM)*, 2012.

[50] P. Traynor, M. Lin, M. Ongtang, V. Rao, T. Jaeger, P. McDaniel, and T. La Porta, "On Cellular Botnets: Measuring the Impact of Malicious Devices on a Cellular Network Core," in *Proceedings of the 16th ACM Conference on Computer and Communications Security (CCS)*, 2009.

[51] M. G. Kang, H. Yin, S. Hanna, S. McCamant, and D. Song, "Emulating Emulation-Resistant Malware," in *Proceedings of the 2nd Workshop on Virtual Machine Security (VMSec)*, 2009.

[52] D. Balzarotti, M. Cova, C. Karlberger, E. Kirda, C. Kruegel, and G. Vigna, "Efficient Detection of Split Personalities in Malware," in *Proceedings of the 17th Annual Network & Distributed System Security Symposium (NDSS)*, 2010.

[53] M. Lindorfer, C. Kolbitsch, and P. Milani Comparetti, "Detecting Environment-Sensitive Malware," in *Proceedings of the 14th International Conference on Recent Advances in Intrusion Detection (RAID)*, 2011.

[54] F. Matenaar and P. Schulz, "Detecting Android Sandboxes," http://dexlabs.org/blog/btdetect, August 2012.

[55] P. Ferrie, "Attacks on More Virtual Machine Emulators," Symantec Research White Paper, Tech. Rep., 2007.

[56] C. Kolbitsch, E. Kirda, and C. Kruegel, "The Power of Procrastination: Detection and Mitigation of Execution-Stalling Malicious Code," in *Proceedigns of the 18th ACM Conference on Computer and Communications Security (CCS)*, 2011.

[57] R. Unucheck, "The most sophisticated Android Trojan," http://www.securelist.com/en/blog/8106/The_most_sophisticated_Android_Trojan, June 2013.

[58] Comodo Malware Analysis Team, "Android OBAD - Technical Analysis Paper," http://www.comodo.com/resources/Android_OBAD_Tech_Reportv3.pdf, 2013.

[59] Z. Ashraf, "DIY: Android Malware Analysis – Taking apart OBAD (part 1)," http://securityintelligence.com/diy-android-malware-analysis-taking-apart-obad-part-1/, Oct 2013.

[60] X. Jiang, "Security Alert: New Sophisticated Android Malware DroidKungFu Found in Alternative Chinese App Markets," http://www.csc.ncsu.edu/faculty/jiang/DroidKungFu.html, 2011.

[61] Symantec Security Response, "First Malicious Use of 'Master Key' Android Vulnerability Discovered," http://www.symantec.com/connect/blogs/first-malicious-use-master-key-android-vulnerability-discovered, July 2013.